\documentclass[11pt]{article}
\usepackage{graphicx}
\usepackage{epsfig}
\usepackage{latexsym}
\usepackage{mathrsfs}
\usepackage{feynmf}
\usepackage{balance}
\usepackage{amsfonts}

\title{Stable solitary waves in Super dense plasmas at external magnetic fields}
\author{Azam Ghaani\thanks{Email: az.ghaani@stu-mail.um.ac.ir} $\,$, 
Kurosh Javidan\thanks{Email: Javidan@um.ac.ir} and Mohsen Sarbishaei \thanks{Email: sarbishei@um.ac.ir}\\ \\
\small{Department of Physics, Ferdowsi University of Mashhad,}\\
\small{91775-1436  Mashhad, Iran} }
\date{}
\begin{document}
\maketitle
\begin{abstract}
propagation of localized waves in a Fermi-Dirac distributed super dense matter at the presence of strong external magnetic fields is studied using the reductive perturbation method. Previous works indicate that localized waves break down in unmagnetized super dense hadronic matter. We have shown that stable solitons can be created in such non-relativistic fluids in the presence of an external magnetic field. Such solitary waves are governed by the Zakharov-Kuznetsov (ZK) equation. Properties of solitonic solutions are studied in media with different values of back ground mass density and strength of magnetic field. 
\end{abstract}

\section*{I. Introduction}

compact astrophysical objects in the context of supernova, white dwarfs, neutron stars, etc are results of a gravitational collapse in stars whose core mass exceed the Chandrasekhar limit. They are the densest observable bodies in our universe and have proven to be ideal test bodies for understanding the behaviour of matter under extreme conditions of high pressures, densities and strong electromagnetic and gravitational fields. During the last decade a great progress is occurred in the observational astrophysics in the direction of studying the properties of such compact objects and specially neutron stars \cite{i1,i2,i3}. 

Recent observations related to anomalous x-ray pulsars and soft gamma-ray repeaters  \cite{i4, i5, i6} also prove the existence of neutron stars with very strong magnetic fields which are known as magnetars \cite{i7, i8, i9}. The magnetic field at the surface of the magnetars may be as strong as $10^{11-12} T$. It is estimated that the strength of interior magnetic field in neutron stars may be as large as $10^{15-16}T$\cite{i10, i11}. A magnetic field of such intensity corresponds to a force of $|e|B \approx  1GeV^2$. It is clear that this interaction can significantly affect the properties of the system. Discoveries of huge magnetic field in neutron stars seem to enforce us to study the effects of the magnetic field in compact stars.

Behaviour of hadronic matters in the presence of external magnetic field can be described using a set of equations which called equations of state (EOS). It may be noted that hadronic matters with different constituents, densities and temperatures are described with different EOS. As the density, temperature and ingredients of sections of compact stars are widely different, one have to use different models of EOS for different sections using the available information (or theoretical estimations) \cite{i12, i13,i14, i15}. The relation between EOS of hadronic matter and compact stars identifications is a bidirectional relation. One can find some constraint on the EOS of hadronic matter using observational information from neutron stars too \cite{i16}. 

The structures of super dense objects are very symmetric. Therefore one has to find information from such objects through perturbation behaviours. propagation of localized defects in spatial distribution of plasma particles density or its energy density due to perturbations has widely investigated in different kinds of plasmas. Evolution of solitary waves in Fermi-Dirac plasmas also has been recently investigated but in unmagnetized environments \cite{i17, i18, i19}. It is interesting to study the effects of magnetic fields on the behaviour of small amplitude localized acoustic waves in super dense objects. 

Motivated by such these cases, theoretical study on the effects of extremely large external magnetic fields on the small amplitude localized waves in Fermi-Dirac distributed dense matters is presented here. This paper is organized as follows: In the next section we review the non-relativistic fluid hydrodynamics. In section III we review the equation of state of the quark gluon plasma according to the MIT bag model. In section IV we combine the hydrodynamic equations with the equations of state using the reductive perturbation method (RPM) and derive the differential equations which govern the time evolution of perturbations at zero temperature. In section V we present an analytical study for wave equations. The last section is devoted to some concluding remarks.

\section*{II. Non relativistic fluid dynamics}

In the framework of non-relativistic fluid dynamics, the Navier-Stokes equation is essentially the simplest equation describing the motion of a fluid which is derived under a quite simple physical assumptions \cite{NS 2}. The non relativistic Navier-Stokes equation is indeed the application of Newton's second law of motion for a fluid. 

The non-linear dynamics of the waves propagating on a baryonic matter is governed by the continuity and non relativistic Navier-Stokes equations as following respectively:

\begin{equation}\label{rho}
\frac{\partial\rho_{B}}{\partial t}+\vec{\nabla}. (\rho _B \vec{v})=0
\end{equation}

\begin{equation}\label{NV}
\Sigma{\vec{F}}=\rho\left(\frac{\partial \vec{v}}{\partial t}+\vec{v} . \vec{\nabla}\vec{v}\right) 
\end{equation}

where $\rho $ and $\rho _B$ are the mass density and baryon density of fluid matter, $v$ is the particle fluid speed and $\vec{F}$ is the sum of the forces acting on the fluid. Many different forces may be imposed on a fluid. Here, we just consider pressure and magnetic forces, so the Naveir-Stokes equation (\ref{NV}) becomes:
\begin{equation}\label{NV1}
\frac{\partial v^i}{\partial t}+v^k \frac{\partial v^i}{\partial x^k}=-\frac{1}{\rho}\left(\frac{\partial p}{\partial x^i} -\rho_{C} B(\vec{v}\times\hat{x})^i\right) 
\end{equation}
where $B$ is an external constant magnetic field in the $x$ direction and $\rho_{C}$ is the charge density corresponding to the charged baryons (or baryons constituents) in the environment.

Due to the existence of the magnetic field, the fluid particles (containing positive or negative charges) exhibit different trajectories and we have to consider the continuity and non relativistic Navier-Stokes equations for each particle as follows separately:

\begin{equation}\label{rhoj}
\frac{\partial\rho_{Bj}}{\partial t}+\vec{\nabla} . (\rho_{Bj} \vec{v}_j)=0
\end{equation}

\begin{equation}\label{NV1j}
\frac{\partial v^i_j}{\partial t}+v^k_j \frac{\partial v^i_j}{\partial x^k}=-\frac{1}{\rho}\left(\frac{\partial p}{\partial x^i} -\rho_{Cj} B(\vec{v}_j\times\hat{x})^i\right) 
\end{equation}
where $j$ index is used to represents each individual particles.

The hydrodynamics basic equations (Continuity equation for the baryon density (\ref{rhoj}) and the Navier-Stokes equation (\ref{NV1j})) are supplemented by the equation of state, which relates fluid energy ($ \varepsilon $) to its pressure ($ p $) and describes the fluid motion starting from a given initial condition. 

\section*{III. The QGP equation of state}

Based on extreme necessary conditions, high densities and/or high temperatures, due to the formation of quark gluon plasma, we can expect to find quark gluon plasma at the core of the compact astrophysical objects. In these objects, nucleons (baryons) are compressed to each other and creating a soup of the free quarks and gluons which is called cold QGP \cite{i20}. It is shown that the viscosity of quark matters is almost zero in low temperatures and under large external magnetic fields \cite{vm}. Therefore one can use hydrodynamic to describe the motion of this fluid \cite{viscose}. As the temperature is very low, we consider the framework of non-relativistic hydrodynamics for the fluid in this step and suppose that the equation of state of  plasma is derived by the MIT bag model.  In order to keeping the non-linearities of the theory, we perform the RPM to combine the equation of state of QGP and Navier-Stokes equation \cite{rpm 1}.

The fundamental idea of the MIT bag model help us to derive the equation of state for the QGP system. It describes the QGP as an ideal gas of non interacting quarks and gluons. Inside the bag, quarks treat as a gas of non interacting quarks, move freely to a first approximation and the interactions with gluons is not taken into account. The effects of confinement in this model interpreted through the bag constant ${\bf{B}}_{bag} $ as the energy needed to create a bag in the QCD vacuum \cite{rpm 1}. The boundary condition of confinement of MIT bag model corresponds to the zero quark mass inside the bag but infinity at the boundary or outside the bag \cite{i22}. Calculated mass for $ u$ and $d$ quarks in dense quark matter are about $ 5 MeV$ and therefore the mass of these quarks are negligible in this situation \cite{bu}. On the other hand the quark effective mass also reduces in strong magnetic fields \cite{me}. The baryon density is given by \cite{ER} : 
\begin{equation}\label{B1}
\rho_{B}=\frac{1}{3}\frac{\gamma_{Q}}{(2\pi)^{3}}\int d^{3}k [n_{\vec{k}} - \bar{n}_{\vec{k}}]
\end{equation}
where $n_{\vec{k}}$ and $\bar{n}_{\vec{k}}$ are quark and anti quark distribution functions which for our problem are given by the Fermi-Dirac distribution function:
\begin{equation}\label{N}
n_{\vec{k}}\equiv n_{\vec{k}}(T)= \frac{1}{1+e^{( k -\frac{1}{3}\mu)/T}}
\end{equation}
and
\begin{equation}\label{Nbar}
\bar{n}_{\vec{k}}\equiv \bar{n}_{\vec{k}}(T)= \frac{1}{1+e^{(k+\frac{1}{3}\mu)/T}}
\end{equation}
where $\mu$ is the baryon chemical potential. with considering the gluon and quark contributions, the energy density and the pressure are given by\cite{i23}: 
\begin{equation}\label{E}
\varepsilon={\bf{B}}_{bag}+\frac{\gamma_{G}}{(2\pi)^{3}}\int d^{3}k \:k \:(e^{k /T}-1)^{-1}+\frac{\gamma_{Q}}{(2\pi)^{3}}\int d^{3}k \:k \:[n_{\vec{k}} + \bar{n}_{\vec{k}}]
\end{equation}
\begin{equation}\label{P}
p= -{\bf{B}}_{bag}+\frac{1}{3}\left( \frac{\gamma_{G}}{(2\pi)^{3}}\int d^{3}k \:k \:(e^{k /T}-1)^{-1}+\frac{\gamma_{Q}}{(2\pi)^{3}}\int d^{3}k \:k \:[n_{\vec{k}} + \bar{n}_{\vec{k}}]\right) 
\end{equation}

This model just consider two flavours of quarks (u,d), so that, the degeneracy factors are $\gamma_{G}=16$ for gluons and $\gamma_{Q}=12$ for quarks. It may be noted that the contribution of gluons in the energy and momentum is zero at the zero temperature.  From the above expressions we can derive that:
\begin{equation}
p=\frac{1}{3}\varepsilon - \frac{4}{3}{\bf{B}}_{bag}
\end{equation}
and the speed of sound, $c_{s}$ is given by:
\begin{equation}
c_{s}^2=\frac{\partial p}{\partial\varepsilon}=\frac{1}{3}
\end{equation}

The quark distribution function of cold QGP at zero temperature is the step function. Such this medium can be found in the core of a dense star which its temperature is close to zero and its baryon density is very high \cite{Non-linear}. At zero temperature the expression for the baryon density (\ref{B1}) becomes:
\begin{equation}\label{B2}
\rho_{B}=\frac{2}{3\pi^{2}}k_{F}^{3}
\end{equation}
where $k_{F}$ is the highest occupied momentum level. Using (\ref{B2}) in (\ref{E}) and (\ref{P}) we rewrite the energy density and pressure as:
\begin{equation}\label{E1}
\varepsilon(\rho_{B})=\left( \frac{3}{2}\right) ^{7/3}\pi^{2/3} \rho_{B}^{4/3} +{\bf{B}}_{bag}
\end{equation}
\begin{equation}\label{P1}
p(\rho_{B})=\frac{1}{3}\left(\frac{3}{2}\right)^{7/3}\pi^{2/3} \rho_{B}^{4/3} -{\bf{B}}_{bag}
\end{equation}
From (\ref{P1}) we have:
\begin{equation}\label{Gr}
\vec{\nabla}p= \frac{4}{9}\left(\frac{3}{2}\right)^{7/3}\pi^{2/3} \rho_{B}^{1/3}\vec{\nabla}\rho_{B}
\end{equation}
In the non relativistic limit $\varepsilon+p\cong\rho$ \cite{viscose,rpm 1} and therefore:
\begin{equation}\label{Nrho}
\rho=\frac{4}{3}\left(\frac{3}{2}\right)^{7/3}\pi^{2/3} \rho_{B}^{4/3}
\end{equation}
the above equations (\ref{Gr}) and (\ref{Nrho}) can be used in the equation (\ref{NV1j}) as the results of the equation of state.

\section*{IV. Non linear wave equation in QGP }

To simplify the problem, we suppose that the anti quarks are negligible in comparison with quarks of the medium. Because of consideration the zero temperature, this assumption does not change the equation of state. Also, we assume that the flavour changing processes are negligible and  thus we can write the continuity equation for each baryon density seperately. The above assumptions are acceptable for cold QGP. Therefore we have:

\begin{equation}\label{RH}
\rho_{Bd} = \alpha \rho_{Bu}
\end{equation}
For the mass density of each quarks we obtain:
\begin{equation}\label{Rm}
\rho_u = \frac{1}{1+\alpha} \rho   \qquad \qquad and  \qquad \qquad   \rho_d = \frac{\alpha}{1+\alpha} \rho
\end{equation}
and the relation between charge density and baryon density for quarks are $\rho_{Cu} = 2 \rho_{Bu}$ and  $\rho_{Cd} = -\rho_{Bd}$. According to the (\ref{rhoj}) and (\ref{NV1j}) the hydrodynamic equations become:
\setlength\arraycolsep{1.pt}
\begin{eqnarray}\label{Tou}
 \frac{\partial\rho_{Bu}}{\partial t} &+&\vec{\nabla} . (\rho_{Bu} \vec{v}_u)=0 \nonumber\\
 \frac{1}{1+\alpha} &\rho& \left( \frac{\partial v^i_u}{\partial t} +v^k_u \frac{\partial v^i_u}{\partial x^k}\right) =-\left(\frac{\partial p}{\partial x^i} - 2\rho_{Bu} B(\vec{v}_u\times\hat{x})^i\right) 
\end{eqnarray}
\begin{eqnarray}\label{Tod}
\frac{\partial\rho_{Bd}}{\partial t}&+&\vec{\nabla} . (\rho_{Bd} \vec{v}_d)=0 \nonumber\\
\frac{\alpha}{1+\alpha} &\rho & \left( \frac{\partial v^i_d}{\partial t}+v^k_d \frac{\partial v^i_d}{\partial x^k}\right) =-\left(\frac{\partial p}{\partial x^i} +\rho_{Bd} B(\vec{v}_d\times\hat{x})^i\right)
\end{eqnarray}

Time evolution of the baryon density in the cold QGP phase can be studied by inserting (\ref{Gr}) and (\ref{Nrho}) into equations (\ref{Tou}) and (\ref{Tod}). The Reductive perturbation Method (RPM) is a technique which is usually used for non-linear wave equations. In this method the non-linear effects, dissipative and dispersive terms are preserved in the wave equations  \cite{Non-linear, rpm 1,rpm 2}.  We expand the equations in powers of a small parameter $\sigma$ and combine these equations to find differential equation(s) which govern the space time evolution of the perturbation in the baryon density. 

We consider perturbations in (3+1) dimensions with Cartesian coordinates. At first we define the dimensionless variables:
\begin{equation}\label{Da}
\hat{\rho_ j}=\frac{\rho_{Bj}}{\rho_{0}}   \quad,\qquad   \hat{v}_{j}=\frac{\vec{v}_{j}}{c_s} 
\end{equation}
where $ j $ represents the species ($j= u , d $) and $\rho_0$ is the background baryon density of the fluid, upon which the perturbation propagates. So we can rewrite (\ref{Tou}) and (\ref{Tod}) using (\ref{Da}) by introducing the following variables $\tau$, $\xi$ and $\phi$ as the stretched coordinates:
\begin{equation}\label{str}
\tau=\sigma^{3/2}t \qquad,\qquad  \xi=\sigma^{1/2}(l_x x+l_z z- V_0 t)\qquad,\qquad   \phi=\sigma^{1/2} y
\end{equation}
where $l_x $ and $l_z$ are the directional cosines of the wave vector  $ \vec{k} $  along the $x$ and $ z $ axes, so that $l^2_x +l^2_z=1$ and  $V_0 $  is an unknown wave phase speed which has to be calculated. It is clear that different kinds of medium constituents have different phase speed. We can now expand the dimensionless baryon density and fluid velocity of components in power series of $\sigma$ as follows:
\begin{eqnarray}\label{Ha}
\hat{\rho_j}&=&1+\sigma\rho_{j1}+\sigma^2\rho_{j2}+...\nonumber \\
\hat{v}_{xj}&=&\sigma v_{xj1}+\sigma^2 v_{xj2}+...\nonumber \\
\hat{v}_{yj}&=&\sigma^2 v_{yj1}+\sigma^3 v_{yj2}+...\\
 \hat{v}_{zj}&=&\sigma^{3/2} v_{zj1}+\sigma^{5/2} v_{zj2}+...\nonumber 
\end{eqnarray}
Note that in an external magnetic field, particles tend to move along the magnetic field freely. We now use the stretched coordinates (\ref{str}), expansions  (\ref{Ha}) and neglecting higher order terms of $\sigma$ in the equations (\ref{Tou}) and (\ref{Tod}).
To the lowest order of $\sigma$ in continuity and momentum equations we can obtain the following results:
\begin{equation}\label{sig}
\rho_{u1}=\frac{l_x c_s }{V_{0u}} v_{xu1} \qquad ,\qquad \rho_{d1}=\frac{l_x c_s }{V_{0d}} v_{xd1} 
\end{equation}
\begin{equation}\label{sig0}
v_{xu1}=\frac{(1+\alpha) l_x }{3 c_s V_{0u}}\rho_{u1}  \qquad ,\qquad v_{xd1}=\frac{(1+\alpha) l_x }{3 c_s \alpha V_{0d}} \rho_{d1}   
\end{equation}
\begin{equation}\label{L}
l_z=0  \qquad  consequently  \qquad l_x=1 
\end{equation}
\begin{equation}\label{sig1}
 v_{zu1} = \frac{(1+\alpha)^{4/3}}{2\rho_0 B} \frac{A}{3c_s }\frac{\partial\rho_{u1}}{\partial\phi} \qquad,\qquad v_{zd1} = -\left( \frac{1+\alpha}{\alpha}\right) ^{4/3}\frac{A}{3c_s \rho_0 B}\frac{\partial\rho_{d1}}{\partial\phi}
\end{equation}
where
\begin{equation}\label{Aa}
A=\frac{4}{3}\left(\frac{3}{2}\right)^{7/3}\pi^{2/3}\rho_0^{4/3}
\end{equation}
The phase speeds $V_{0j}$ can be calculated from (\ref{sig}) and (\ref{sig0}) as follows:
\begin{equation}\label{v}
V_{0u}=\sqrt{\frac{1+\alpha}{3}} \qquad,\qquad V_{0d}=\sqrt{\frac{1+\alpha}{3\alpha}}
\end{equation}
From the terms of the order $\sigma^{3/2} $ we obtain:
\begin{equation}\label{yz}
v_{yu1}=(1+\alpha)^{1/3} \frac{AV_{0u}}{2\rho_0 B}\frac{\partial v_{zu1}}{\partial\xi} \qquad ,\qquad v_{yd1}=-\left( \frac{1+\alpha}{\alpha}\right)^{4/3} \frac{AV_{0d}}{\rho_0 B}\frac{\partial v_{zd1}}{\partial\xi}
\end{equation}
which, after the derivation with respect to $\phi $, become:
\begin{equation}\label{yz1}
\frac{\partial v_{yu1}}{\partial\phi}=(1+\alpha)^{1/3} \frac{AV_{0u}}{2\rho_0 B} \frac{\partial ^2 v_{zu1}}{\partial\xi \partial\phi} \quad,\quad \frac{\partial v_{yd1}}{\partial\phi}=-\left( \frac{1+\alpha}{\alpha}\right)^{4/3} \frac{AV_{0d}}{\rho_0 B} \frac{\partial ^2 v_{zd1}}{\partial\xi \partial\phi}
\end{equation}
Also from the terms proportional to $\sigma^{2} $ and using equation (\ref{sig}) and (\ref{v}) we find:
\begin{equation}\label{r1}
\frac{\partial\rho_{u1}}{\partial\tau}+ \sqrt{\frac{1+\alpha}{3}}\rho_{u1}\frac{\partial\rho_{u1}}{\partial\xi}+\frac{1}{2} c_s \frac{\partial v_{yu1}}{\partial\phi}=0
\end{equation}
\begin{equation}\label{r11}
\frac{\partial\rho_{d1}}{\partial\tau}+\sqrt{\frac{1+\alpha}{3\alpha}}\rho_{d1}\frac{\partial\rho_{d1}}{\partial\xi}+\frac{1}{2} c_s \frac{\partial v_{yd1}}{\partial\phi}=0
\end{equation}
Equations (\ref{r1}) and (\ref{r11}) can be written as the following forms using  (\ref{yz1}) and (\ref{sig1}):
\begin{equation}\label{r2}
\frac{\partial\rho_{u1}}{\partial\tau}+ \sqrt{\frac{1+\alpha}{3}}\rho_{u1}\frac{\partial\rho_{u1}}{\partial\xi}+\frac{A^2}{6\sqrt{3}(2\rho_0 B)^2}(\alpha+1)^{13/6} \frac{\partial ^3 \rho_{u1}}{\partial \phi ^2\:\partial\xi}=0
\end{equation}
\begin{equation}\label{r22}
\frac{\partial\rho_{d1}}{\partial\tau}+\sqrt{\frac{1+\alpha}{3\alpha}}\rho_{d1}\frac{\partial\rho_{d1}}{\partial\xi}+\frac{A^2}{6\sqrt{3}(\rho_0 B)^2}\left(\frac{1+\alpha}{\alpha}\right)^{13/6} \frac{\partial ^3 \rho_{d1}}{\partial \phi ^2\:\partial\xi}=0
\end{equation}
Equations (\ref{r2}) and (\ref{r22}) are known as the Zakharov-Kuznetsov (ZK) equation which is another alternative version of non-linear model describing two dimensional modulation of the Korteweg-de Vries (KdV) solitons \cite{Zkm} when the magnetic field is directed along the $x$ axis. Fortunately there exist exact localized wave solutions for the ZK equation.

The above equation in the Cartesian coordinates becomes:
\begin{equation}\label{fin}
\frac{\partial \hat{\rho}_{u1}}{\partial t}+\sqrt{\frac{1+\alpha}{3}} \frac{\partial\hat{\rho}_{u1}}{\partial x}+ \sqrt{\frac{1+\alpha}{3}} \hat{\rho}_{u1}\frac{\partial\hat{\rho}_{u1}}{\partial x}+\frac{A^2}{6\sqrt{3}(2\rho_0 B)^2}(1+\alpha)^{13/6}\frac{\partial ^3 \hat{\rho}_{u1}}{\partial x  \: \partial y^2}=0
\end{equation}
\begin{equation}\label{fin1}
\frac{\partial \hat{\rho}_{d1}}{\partial t}+ \sqrt{\frac{1+\alpha}{3\alpha}} \frac{\partial\hat{\rho}_{d1}}{\partial x}+\sqrt{\frac{1+\alpha}{3\alpha}} \hat{\rho}_{d1}\frac{\partial\hat{\rho}_{d1}}{\partial x}+\frac{A^2}{6\sqrt{3}(\rho_0 B)^2}\left(\frac{1+\alpha}{\alpha}\right)^{13/6}\frac{\partial ^3 \hat{\rho}_{d1}}{\partial x  \: \partial y^2}=0
\end{equation}
with $\hat{\rho}_{j1}\equiv\sigma\rho_{j1}$.

\section*{V. discussion}

The ZK equation is one of the best known two-dimensional generalizations of the KdV equation \cite{Zkm1}. For a magnetic field which is directed along the $\xi$-axis in the $(\xi,\phi, \tau) $ space, the ZK equation has the general form of \cite{Zkm}:
\begin{equation}\label{ZK}
u_{\tau}+a \left( u\right) ^{2}_{\xi}+\left( bu_{\xi\xi} +ku_{\phi\phi}\right)_{\xi} =0
\end{equation}
where $a$, $b$ and $k$ are constants. An exact solution for a given non-linear partial differential equation can be found using homogeneous balance (HB) method. According to this method, the travelling wave solution of the ZK equation is obtained as follows:
\begin{equation}\label{ans}
u(\xi,\phi,\tau)=\frac{1}{2a}\left( d - 8(b+k)\alpha\beta - 12(b+k) tanh^{2}(\xi+\phi-d\tau)\right) 
\end{equation}
where $ d $ is a constant and $\alpha\beta=-1$. After changing variables to the $ (x,y,t) $ space, we have:
\begin{equation}\label{ZKch}
\hat{u}_{t}+V_0 \hat{u}_{x}+a  \left( \hat{u}\right) ^{2}_{x}+ \left( b\hat{u}_{xx} +k\hat{u}_{yy}\right)_{x} =0
\end{equation}
where $\hat{u}\equiv \sigma u$.  Now one can apply the following set of new coordinates to the ZK solution (\ref{ans}):
\[ \tau=t  \qquad ,\qquad \xi=x  - V_0 t \qquad and \qquad \phi=y \]
and the exact localized solution of (\ref{ZKch}) becomes:
\begin{equation}\label{ansch}
\hat{u}(x,y,t)=\frac{1}{2a}\left( d + 8(b+k)- 12(b+k) tanh^{2}(x+y -(d+V_0)t)\right) 
\end{equation}
According to the equations (\ref{fin}) and using the relation (\ref{Aa}), the constant coefficients are:
\[a=\frac{1}{2}\sqrt{\frac{1+\alpha}{3}}     \qquad ,\qquad b=0  \qquad and \qquad k=\frac{(\frac{4}{3}\left(\frac{3}{2}\right)^{7/3}\pi^{2/3}\rho_0^{4/3})^{2}}{6\sqrt{3}(2\rho_0 B)^{2}} (1+\alpha)^{13/6} \]
and for the equation (\ref{fin1}) we obtain:
\[a=\frac{1}{2}\sqrt{\frac{1+\alpha}{3\alpha}}     \qquad ,\qquad b=0  \qquad and \qquad k=\frac{(\frac{4}{3}\left(\frac{3}{2}\right)^{7/3}\pi^{2/3}\rho_0^{4/3})^{2}}{6\sqrt{3}(\rho_0 B)^{2}} \left( \frac{1+\alpha}{3\alpha}\right) ^{13/6} \]
Finally, the solution of (\ref{fin}) and (\ref{fin1}), as a perturbation in the baryon density at zero temperature, is obtained as:
\setlength\arraycolsep{1.pt}
\begin{eqnarray}\label{R}
\hat{\rho}_{u1}(x,y,t)&=& \sqrt{\frac{3}{1+\alpha}} \left( d +8\frac{(\frac{4}{3}\left(\frac{3}{2}\right)^{7/3}\pi^{2/3}\rho_0^{4/3})^{2}}{6\sqrt{3}(2\rho_0 B)^{2}} (1+\alpha)^{13/6}\right) \nonumber\\
 - \sqrt{\frac{3}{1+\alpha}}&&\left(12\frac{(\frac{4}{3}\left(\frac{3}{2}\right)^{7/3}\pi^{2/3}\rho_0^{4/3})^{2}}{6\sqrt{3}(2\rho_0 B)^{2}} (1+\alpha)^{13/6} tanh^{2}\left( x+y -(d+\sqrt{\frac{1+\alpha}{3}})t\right) \right) \nonumber\\
\end{eqnarray}
and
\begin{eqnarray}\label{R1}
\hat{\rho}_{d1}(x,y,t)&=& \sqrt{\frac{3\alpha}{1+\alpha}} \left( d+8\frac{(\frac{4}{3}\left(\frac{3}{2}\right)^{7/3}\pi^{2/3}\rho_0^{4/3})^{2}}{6\sqrt{3}(\rho_0 B)^{2}} \left( \frac{1+\alpha}{3\alpha}\right) ^{13/6}\right) \nonumber\\
-\sqrt{\frac{3\alpha}{1+\alpha}}&&\left(12\frac{(\frac{4}{3}\left(\frac{3}{2}\right)^{7/3}\pi^{2/3}\rho_0^{4/3})^{2}}{6\sqrt{3}(\rho_0 B)^{2}} \left( \frac{1+\alpha}{3\alpha}\right) ^{13/6} tanh^{2}\left( x+y -(d+\sqrt{\frac{1+\alpha}{3\alpha}})t\right) \right)\nonumber\\ 
\end{eqnarray}
where $ \alpha $ is a free parameter which can be determined from the state of matter. In our problem $\alpha$ is calculated by charge density continuity equation:
\begin{equation}\label{c}
\frac{\partial\rho_{C}}{\partial t}+\vec{\nabla}. \vec{J}_C=0
\end{equation}
where $\vec{J}_C = \rho_{Cu} \vec{v}_u + \rho_{Cd} \vec{v}_d$. From the Maxwell equation with regardless of the time derivative of the electric field, $\vec{\nabla}. \vec{J}_C=0$ and thus  $\frac{\partial\rho_{C}}{\partial t}=0$. Therefore $\rho_c=\rho_{cu}+\rho_{cd}$ is constant. For neutron stars electric charge density is zero. From that the electric charge of $u$ quark is $-2$ times of the electric charge of $d$ quark we can write $\rho_{Bd}=2\rho_{Bu}$ and therefore $\alpha=2$. 

Based on the latest available information on neutron stars, we have chosen $\rho_0 = 0.2  fm^{-3} $  ,  $ B=10^{11}T \simeq 6.67 MeV^{2}/e $ and $ c_{s}^{2} = 1/3 $ \cite{B, rho}, as typical values for the core of NS in our simulations. Figure 1 presents the time evolution of $u$ quark and $d$ quark densities. This figure shows that pulses propagate without distortion in a direction which makes the angle of $\frac{\pi}{4}$ with the magnetic field direction. It is an interesting result. It may be noted that such waves are able to reach the border of the medium and create measurable effects out of the region in a homogeneous background. This figure also demonstrates that the wave phase speed (amplitude) of $u$ quark density perturbation is greater (smaller) than the phase velocity (amplitude) of $d$ quark density perturbation. The actual value of the $\rho_1$ is small and the figures has been plotted out of scale.
\begin{figure}[htp]\label{fig1}
\centerline{\begin{tabular}{cc}
\includegraphics[width=11 cm, height=8 cm]{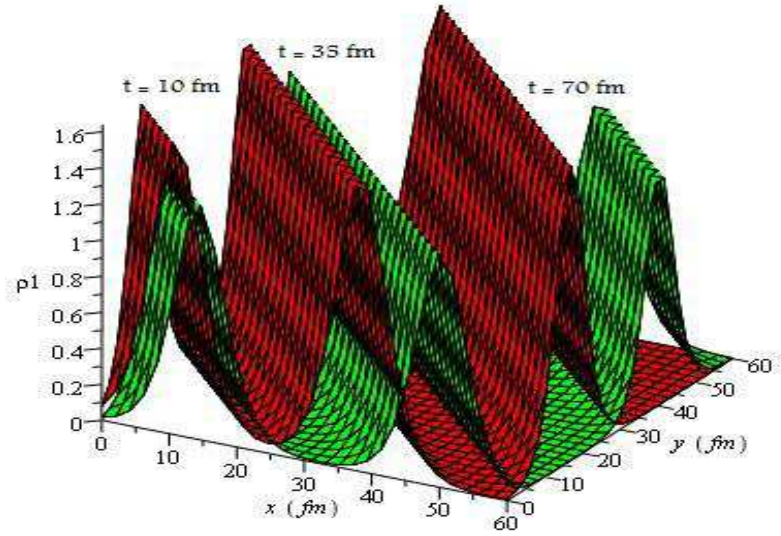}
\end{tabular}}
 \caption{\footnotesize
 The time evolution of baryons density perturbation in the present of a magnetic field. The green plot is for $\rho_{u1}$ and  red plot is for  $\rho_{d1}$  }
\end{figure}

Equation (\ref{R}) shows that the amplitude of the wave is a function of background density $\rho_0 $ , magnetic field $ B $ and $ \alpha $. Figure 2 demonstrates created localized waves in media with different values of relative background baryon density $\rho_{0}$. This figures shows that the soliton amplitude increases with increasing values of $\rho_{0}$ while its ground level respect to the background decreases. 

Figure 3 presents soliton profiles created in media with different values of relative magnetic field as functions of $x$. The figure shows that the soliton amplitude and also its ground level decreases as the magnetic field increases. Therefore we can conclude that strong magnetic fields kill the localized perturbations. But it may be noted that stable solitons in such media are established due to the presence of the magnetic field.     

\begin{figure}[htp]\label{fig2}
\centerline{\begin{tabular}{cc}
\includegraphics[width=11 cm, height=8 cm]{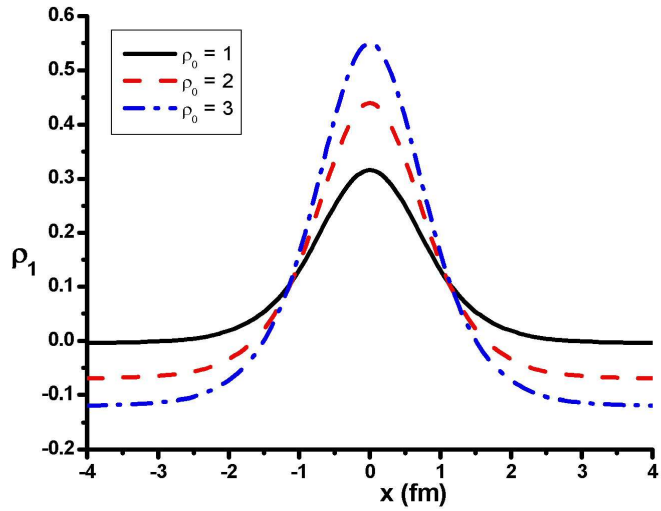}
\end{tabular}}
 \caption{\footnotesize
 Soliton profiles as functions of $x$ with $y=0$ and $t=0$ and different values of background baryon density $\rho_{0}$. }
\end{figure}

\begin{figure}[htp]\label{fig3}
\centerline{\begin{tabular}{cc}
\includegraphics[width=11 cm, height=8 cm]{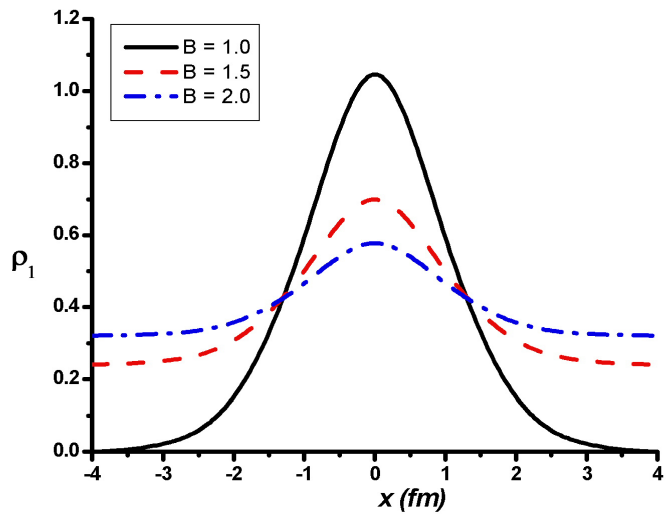}
\end{tabular}}
 \caption{\footnotesize
 Soliton profiles as functions of $x$ with $y=0$ and $t=0$ and different values of background Magnetic fields $B$. }
\end{figure}

\section*{VI. Conclusions and remarks}

The study of the propagation of localized waves in uniform hadronic matter and QGP phase is a very important problem, because the core of the compact astrophysical objects like neutron stars may contain de confined quark matter at high baryon densities and low temperatures. It is estimated that the interior magnetic field in the core and at the surface of neutron stars may be as large as $ 10^{11-16} T $. So that, it is important to study the influence of the magnetic field on the behaviour of hadronic matter in neutron stars. In this work we have studied the effect of magnetic field on the wave propagation in a cold QGP. 

Recent investigations have been shown that unstable but long lasting localized breaking waves can be created in unmagnetized super dense hadronic matter and QGP. We have shown that stable solitonic profiles can be created in such these media in the presence of magnetic fields which are solutions of the ZK equation. Moving solitons are able to reach the borders of the medium and create measurable effects in a uniform background. Solitons move in a direction with the angle of $\frac{\pi}{4}$ respect to the magnetic field direction. Soliton amplitude increases as the background  baryon density (i.e. back ground mass density) increases while it decreases with an increasing value of magnetic field.  

There are many open questions in this situation which needs more investigations. What is the behaviour of localized solutions in QGP (or hadronic matters) at finite temperature? We need at first a suitable equation of state for such this medium. What are the effects of medium viscosity? What are the effects of non Fermi-Dirac distributions? It is clear that in non-zero temperature case matters may have different distribution functions. Mass (Baryonic) density and/or magnetic field generally is not constant in all the region of the medium under investigation. What are the effects of space dependent mass density and magnetic field?


\begin{thebibliography}{50}
\bibitem{i1} . D. Barret, J.-F. Olive, and M. C. Miller, Mon. Not. R. Astron. Soc. 361, 855 (2005).
\bibitem{i2} F. Ozel, Nature 441, 1115 (2006).
\bibitem{i3}  J. M. Lattimer and M. Prakash, Phys. Rep. 442, 109 (2007).
\bibitem{i4} C. Kouvellioton, Nature 393, 235 (1998).
\bibitem{i5} K. Hurley et al Astrophys. J 510, L111 (1999).
\bibitem{i6} S. Mareghetti and L. Stella, Astrophys. J. 442 L17 (1995). J. vanParadijs, Astrophys. J. 513 464 (1999).
\bibitem{i7} R. C. Duncan and C. Thompson, Astronphys. J. 392, L9 (1992).
\bibitem{i8} V. V. Usov, Nature 357, 472 (1992).
\bibitem{i9} B. Paczy´nski, Acta Astron. 42, 145 (1992).
\bibitem{i10} R. C. Duncan and C. Thompson, Astrophys. J. 392, L9 (1992).
\bibitem{i11} S. L. Shapiro and S. A. Teukolsky,Black holes, white dwarfs and neutron stars, Wiley-interscience New York, 1983.
\bibitem{i12} Ken’ichiro Nakazato, Kohsuke Sumiyoshi, Shoichi Yamada: Phys.Rev.D 77, 103006 (2008).
\bibitem{i13} Tsuyoshi Miyatsu, Myung-Ki Cheoun, Koichi Saito: JPS Conf. Proc. 1, 013080 (2014).
\bibitem{i14} Tsuyoshi Miyatsu, Sachiko Yamamuro, Ken’ichiro Nakazato: Astrophys.J. 777, 4 (2013).
\bibitem{i15} S. Gandolfi, J. Carlson, S. Reddy, A. W. Steiner, R. B. Wiringa: EPJA 50, 10 (2014).
\bibitem{i16} Hell Thomas, Bernhard Roettgers, Wolfram Weise: Conference Proceedings for INPC, arXiv:1307.4582 (2013).

\bibitem{i17} Ata-ur Rahman S. Ali: Astrophys Space Sci 351, 165-172 (2014).
\bibitem{i18} I. Zeba, W. M. Moslem and P. K. Shukla: The Astrophysical Journal 750, 72 (2012).
\bibitem{i19} S. Mahmood, Safeer Sadiq and Q. Haque: Physics of Plasmas 20, 122305 (2013).
\bibitem{NS 2} Hardi Peter and Rolf Schlichenmaier :"Introduction to Hydrodynamics", Freiburg, (2005).
\bibitem{i20} J. C. Collins and M. J. Perry, Phys. Rev. Lett. {\bf{34}}, 1353 (1975).
\bibitem{vm} Seung-il Nam1 and Chung-Wen Kao, PHYSICAL REVIEW D 87, 114003 (2013) 
\bibitem{viscose} D. A. Fogaca, F. S. Navarra, and L. G. Ferreira Filho, Phys. Rev. C {\bf{88}}, 025208 (2013); arXiv:1305.0798 [nucl-th].
\bibitem{rpm 1}D. A. Fogaca, F. S. Navarra, L. G. Ferreira Filho, arXiv:1212.6932 [nucl-th], (2012).
\bibitem{i22} Yun Zhang and Ru-Keng Su, arXiv:nucl-th/0203007v1, (2002).
\bibitem{bu} M. Buballa, Physics Reports 407, 205–376 (2005)
\bibitem {me} D.P. Menezes, M. Benghi Pinto, S.S. Avancini, A. P´erez Mart´ınez, and C. Providˆencia, Phys.Rev.C 79 :035807 (2009)
\bibitem{ER} Michel le Bellac :"Thermal Field Theory", Cambrige Uneversity Press, (1996).
\bibitem{i23} B.D. Serot and J.D. Walecka, Advances in Nuclear Physics 16, 1 (1986).
\bibitem{Non-linear} D. A. Fogaca, L. G. Ferreira Filho, and F. S. Navarra, Phys. Rev. C {\bf{81}}, 055211 (2010); arXiv:0908.4215 [nucl-th].
\bibitem{rpm 2}H. Washimi and T. Taniuti, Phys. Rev. Lett. {\bf{17}}, 996 (1966).
\bibitem{Zkm1} A.M.Wazwaz, Communications in Nonlinear Science and Numerical Simulation {\bf{10}}, (2005).
\bibitem{Zkm} Mohammed Khalfallah, An. St. Univ. Ovidius Constanta {\bf{15(2)}}, (2007). 
\bibitem{B} Andreas Reisenegger, arXiv:1305.2542v1 [astro-ph.SR], (2013).
\bibitem{rho} G.B.Alaverdyan, A.R.Harutyunyan, Yu.L.Vartanyan, Astrophysics 46 (2003); arXiv:astro-ph/0409602v1.



\bibitem{ent} L. Landau and E. Lifchitz :"Fluid Mechanics", Pergamon Press, Oxford, (1987).
\bibitem{NS 1} S. Weinberg :"Gravitation and Cosmology", New York: Wiley, (1972).





\end{thebibliography}
\end{document}